%% file: skyrmion.tex
\documentclass[12pt]{article}
\usepackage[dvips]{graphicx}
\usepackage{latexsym}
\usepackage{amssymb}
\input{papers_macros.tex}
\newcounter{mnotecount}[section]

\renewcommand{\themnotecount}{\thesection.\arabic{mnotecount}}

\newcommand{\mnote}[1]
{\protect{\stepcounter{mnotecount}}$^{\mbox{\footnotesize
$
\bullet$\themnotecount}}$ \marginpar{
\raggedright\tiny\em
$\!\!\!\!\!\!\,\bullet$\themnotecount: #1} }

\begin{document}
\title{\vskip -70pt
\begin{flushright}
\end{flushright}
\vskip 10pt
{\bf Skyrmions from gravitational instantons
\vskip 10pt}}
\author{Maciej Dunajski\thanks{Email: M.Dunajski@damtp.cam.ac.uk}
\\
Department of Applied Mathematics and Theoretical Physics,\\
University of Cambridge,\\
Wilberforce Road, Cambridge CB3 0WA, UK. }
\date{}
\maketitle

\begin{abstract}
We propose a construction of Skyrme fields from holonomy 
of the spin connection of 
gravitational instantons. The procedure is implemented for Atiyah--Hitchin and Taub--NUT instantons.

The skyrmion resulting from the Taub--NUT 
is given explicitly on the space of orbits of a left translation inside the whole isometry group. The domain of the Taub--NUT 
skyrmion is a trivial circle bundle over the Poincare disc. The position of the skyrmion depends on the Taub--NUT mass parameter,
and its topological charge is equal to two.
\end{abstract}
\section{Introduction}

The Standard Model of elementary particles is a quantum gauge theory 
with a non--abelian gauge group $SU(3)\times SU(2) \times U(1)$. 
This model is at some level fundamental, and provides a complete
field theory of interacting quarks. Thus in principle
it should describe  protons and neutrons. On the other hand 
the Lagrangian underlying the Standard Model leads to non-linear field 
equations, which has so far made it impossible to obtain
exact results about the bound states describing the particles of the theory.

An alternative is to look for theories which ignore the internal structure
of particles, and instead give an effective, low energy description of baryons.
The Skyrme model \cite{Skyrme_paper} is an example of such theory, where 
baryons arise as solitons. The topological degree of these solitons is 
identified with the baryon number in a mechanism which naturally leads to a
topological baryon number conservation.

In a recent paper \cite{AMS11} Atiyah, Manton and Schroers (AMS) 
proposed a far reaching generalisation of the Skyrme model which 
involves several topological invariants, and aims to give a geometrical and
topological interpretations to the electric charge, the baryon number 
and the lepton number.
In the AMS model static particles are described in terms  of gravitational 
instantons - Riemannian four manifolds which satisfy the 
Einstein equations and whose  curvature is 
concentrated in a finite region of a space-time \cite{GH}.

The electrically charged particles correspond to non--compact asymptotically 
locally flat (ALF) instantons $(M, g)$  -- complete  four--dimensional 
Riemannian manifolds which
solve the Einstein equations (possibly with cosmological constant)
and approach $S^1$ bundle over $S^2$  at infinity.
The first Chern class of the  asymptotic $U(1)$ fibration gives the electric charge. Neutral particles
correspond to compact instantons. In all cases the baryon number 
has a topological origin and is identified with the signature of $M$.

The AMS model is inspired by 
the Atiyah--Manton approach \cite {AM89} to the Skyrme model of baryons,
where a static Skyrme field $U:\R^3\rightarrow SU(2)$ with 
the boundary condition
$U(\bf{x})\rightarrow {\bf 1} $ as $|{\bf x}|\rightarrow \infty$
arises from a holonomy of a Yang--Mills instanton on $\R^4$ along one of the 
directions. Thus the physical three--space $\R^3$ is regarded
as the space of orbits of a one--parameter group  of conformal isometries
of $\R^4$. AMS use this as a motivation for their model, but in the AMS
approach the three--space is 
(for the electrically charged particles) the base space of 
asymptotic  circle fibrations.

The idea behind the present paper is to use the AMS model as a motivation 
for relating particles to gravitational instantons, but then to proceed
in a way analogous to the Atiyah--Manton construction to recover a skyrmion
from a gravitational instanton. Thus, in our case, 
the
three--space ${\mathcal B}$ will  arise as a quotient of $M$ by a certain 
$S^1$ 
action. The $SU(2)$ instanton holonomy will be replaced by
a holonomy of a spin connection on $\spp_+$, where $TM\otimes \C\cong
\spp_+\otimes\spp_-$, and $\spp_\pm$ are rank two complex vector bundles
over $M$ (see e. g. \cite{Dunajski_book}).

In the next Section we shall reinterpret the $\mathfrak{su}(2)$ spin connection
on $\spp_+$ as the potential for a self--dual Yang--Mills field 
on the gravitational instanton background. In Section \ref{section3}
we shall compute the holonomy of this potential along the orbits
of an $SO(2)$ left--translation inside the whole isometry group of
$(M, g)$, using the Atiyah--Hitchin and Taub--NUT instantons as examples.
This will give rise to a skyrmion on the space of orbits ${\mathcal B}$
of $SO(2)$ in $M$. We shall find the expression for the topological charge density, and
compute this charge for the Taub--NUT skyrmion. In Section \ref{section4} we shall construct
the Riemannian metric $h_{\mathcal B}$ on the three--dimensional domain 
${\mathcal B}$ of the skyrmion.
In the case of Taub--NUT skyrmion this metric is complete
and describes  a trivial circle fibration over the upper half--plane, 
with circular fibres of non--constant radius:
\[
h_{\mathcal B}=g_{\HH^2}+R^2d\psi^2,\quad\mbox{where}\quad 
R^2=\frac{1}{(\mu r+1)^2\sin{\theta}^2+\cos{\theta}^2}.
\]
Here the constant parameter $\mu$ is the inverse mass in the Taub--NUT space, 
and $y=r\sin{\theta}>0, x=r\cos{\theta}$ are coordinates on $\HH^2$ with the
hyperbolic metric 
$g_{\HH^2}=y^{-2}(dx^2+dy^2)$. In these coordinates the skyrmion is given 
by
\be
\label{skyrmion_introduction}
U=\exp{\Big(i\pi\Big(\frac{r\mu}{r\mu+1}\sin{\theta}\;(\cos\psi\; \tau_1
+\sin\psi\; \tau_2)
-\frac{r\mu(r\mu+2)}{(r\mu+1)^2}\cos{\theta}\;\tau_3 \Big )\Big)}.
\ee
The skyrmion (\ref{skyrmion_introduction}) is localised on the imaginary axis,
around the point $(0,5/(4\mu))$. We should again, at this point, 
emphasise the difference between our construction and the AMS approach.
The `physical' three--space in \cite{AMS11} admits an isometric
$SO(3)$ action, whereas the three--dimensional Riemannian 
manifold $({\mathcal B}, h_{\mathcal B})$ which supports
the skyrmion admits only one isometry in the Taub--NUT case. This is 
because the generator of the left translation used in the construction of the quotient belongs to 
a two--dimensional abelian subalgebra inside the full Lie algebra
of the isometry group $U(2)$. Thus one Killing vector of the Taub--NUT space 
descends down to the quotient. In the Atiyah--Hitchin case the isometry
algebra $SO(3)$ does not contain two--dimensional abelian sub-algebras, and the quotient
space ${\mathcal B}$ does not admit any Killing vectors, or conformal 
Killing vectors.

\section{Spin connection as gauge potential}
\label{section2}
Properties of a single particle are invariant with respect to
ordinary rotations in three-space. Thus the corresponding
instanton should admit $SO(3)$ or its double cover $SU(2)$ as the group of isometries,
i. e. the metric should take the form
\be
\label{metric}
g=f^2 dr^2+(a_1 \eta_1)^2+(a_2\eta_2)^2+(a_3\eta_3)^2,
\ee
where $\eta_i$ are the left invariant one--forms on $SU(2)$
such that 
\[
d\eta_1=\eta_2\wedge\eta_3, \quad d\eta_2=\eta_3\wedge\eta_1,
\quad \quad d\eta_3=\eta_1\wedge\eta_2
\]
and
$(a_1, a_2, a_3, f)$ are functions of $r$. 
There is no loss of generality in this 
diagonal ansatz, as the induced metric can  always be diagonalised on a
surface of constant $r$, and then the Einstein equations imply  \cite{tod_cohomogeneity} that
the non--diagonal components are fixed (to zero) in the `evolution' in $r$.
The diffeomorphism freedom can be used to set $f=-a_2/r$. 

In the AMS setup the Atiyah--Hitchin manifold \cite{AH} 
is a model for the proton and the self--dual
Taub NUT manifold corresponds to the electron.
Although the spin connection $\gamma_-$ of (\ref{metric}) does not 
vanish in the invariant
frame (\ref{metric}),  the curvature of $(\spp_-, \gamma_-)$ is zero as both AH and Taub--NUT metrics have self--dual Riemann curvature. Thus we shall
consider the connection $\gamma=\gamma_+$ on  $\spp_+$. It is best calculated
using the self--dual two--forms \cite{Dunajski_book}
\begin{eqnarray}
\label{two_forms}
\Sigma_i&=&{\bf e}_0\wedge {\bf e}_i+\frac{1}{2}\varepsilon_{ijk} {\bf e}_j\wedge {\bf e}_k,
\quad \mbox{where}\quad i, j, k =1, \dots, 3\quad\mbox{and}\\
{\bf e}_0&=&fdr,\quad {\bf e}_1=a_1\eta_1,\quad {\bf e}_2=a_2\eta_2,\quad 
{\bf e}_3=a_3\eta_3.\nonumber
\end{eqnarray}
The spin connection coefficients $\gamma_{ij}$ are skew--symmetric and
are determined from the relations 
$
d\Sigma_i+\gamma_{ij} \wedge \Sigma_j=0.
$
We find 
\be
\label{connection}
P_1=f_1(r)\eta_1, \quad P_2=f_2(r)\eta_2, \quad P_3=f_3(r)\eta_3,\quad\mbox{and}\;\;  \gamma_{ij}=\varepsilon_{ijk} P_k,
\ee
where the functions $f_i(r)$ depend on the coefficients $a_i(r)$ and their 
derivatives. 
\begin{itemize}
\item
The  Taub--NUT metric is a unique non--flat 
complete self--dual Einstein metric with isometric $SU(2)$ action
such that the generic orbit is three--dimensional, and the $SU(2)$ action
rotates the anti--self--dual two--forms. In this case
\[
a_1=a_2=r\sqrt{\epsilon+\frac{m}{r}}, \quad 
a_3=m\sqrt{\epsilon+\frac{m}{r}}^{-1},
\]
where $\epsilon$ and $m$ are constants.
At $r=0$ the three-sphere of constant $r$ collapses to a point -- 
an example of a NUT singularity. The SD spin connection coefficients\footnote{For comparison, computing the ASD connection
on $\spp_-$ would also give (\ref{connection}), but this time
with $f_1=f_2=f_3=1$. The Maurer--Cartan equations on $SU(2)$ then imply
that the curvature of this connection vanishes, and so the metric
has self--dual Riemannian curvature.} give the Pope--Yuille instanton \cite{PY}
(see \cite{cherkis} for a discussion on more general Yang-Mills instantons on self-dual
ALF spaces)
\be
\label{taub_nut}
f_1=f_2=-\frac{r\epsilon}{r\epsilon+m}, 
\quad f_3=\frac{r\epsilon(r\epsilon+2m)}{(r\epsilon+m)^2}.
\ee
\item 

The Atiyah--Hitchin  metric is a unique (up to taking a double covering)
complete self--dual Einstein metric with isometric $SO(3)$ 
(rather than $SU(2)$) action
such that the generic orbit is three--dimensional, and the action
rotates the anti--self--dual two--forms \cite{AH}. It is a metric on a moduli space of 2-monopoles with fixed centre. The $SO(3)$ isometric action 
can be traced back to the 2-monopole configuration, where the rotation group
acts on the pair of unoriented spectral lines, and $r$ is (a function of)
an angle between these lines. The coordinate $r$ parametrises
the orbits of $SO(3)$ in the moduli space. In this paper we shall use the double cover
of the moduli space of centered 2-monopoles, and, following \cite{AMS11}, still call it
the Atiyah--Hitchin (or AH) manifold.

In the case of the Atiyah--Hitchin metric we shall only need the 
asymptotic 
formulae. The coordinate $r$ ranges between $\pi$ and $\infty$, and at $r=\pi$ the
$SO(3)$ orbits collapses to a two--sphere (a bolt). For large $r$
\[
a_1=a_2=r\sqrt{1-\frac{2}{r}} +O(e^{-r}), 
\quad a_3=-2\sqrt{1-\frac{2}{r}}^{-1} +O(e^{-r}),
\]
which leads to the asymptotic expressions
\be
\label{AHlarge_r}
f_1=f_2=-\frac{r}{r-2}, \quad f_3=\frac{r(r-4)}{(r-2)^2}.
\ee
For  $r$ close to $\pi$ we have
\begin{eqnarray*}
a_1&=&2(r-\pi)+ O((r-\pi)^2) ,\quad a_2=\pi+\frac{1}{2}(r-\pi)
+O((r-\pi)^2),\\
a_3&=&-\pi+\frac{1}{2}(r-\pi) + O((r-\pi)^2),  
\end{eqnarray*}
which gives
\be
\label{AHsmall_r}
f_1=\frac{\pi-r}{\pi}-3, \quad f_2=\frac{\pi-r}{\pi},\quad 
f_3=\frac{r-\pi}{\pi+r}.
\ee
\end{itemize}
To make contact with the `skyrmions from instantons' ansatz 
of \cite{AM89} we need to reinterpret the self--dual spin connection $\gamma$
as $\mathfrak{su}(2)$--valued gauge field $A$. 
This is done \cite{charap, kor} by setting 
\be
\label{gauge_field}
A=P_1\otimes {\bf t}_1+P_2\otimes {\bf t}_2+P_3\otimes {\bf t}_3,
\ee
where the matrices ${\bf t}_i$ generate the Lie algebra
$\mathfrak{su}(2)$ with the commutation relations 
$[{\bf t}_i, {\bf t}_j]=-(1/2)\varepsilon_{ijk} {\bf t}_k$, and the 
one--forms $P_j$ are given by (\ref{connection}).
\vskip 5pt
Topology of the Yang--Mills field is determined by the topology of the 
gravitational instanton (see \cite{dunajski_2} for a related construction where
topology of an abelian vortex is determined by the topology of the underlying background surface). 
The topological charge of the Yang--Mills
instanton is in general fractional, despite the action being finite.
The relation between the Yang--Mills, and 
Einstein curvatures is easily expressed
using the $SO(3)$ representation spaces:
\begin{eqnarray*}
A&=&P_i \otimes {\bf t}_i=\frac{1}{2}\varepsilon_{ijk}\gamma_{jk} \otimes 
{\bf t}_i\\
F&=& dA+A\wedge A=\frac{1}{2}\varepsilon_{ijk}R_{jk}\otimes {\bf t}_i, 
\quad\mbox{where}\;\;R_{ij}=d\gamma_{ij}
+\gamma_{ik}\wedge\gamma_{kj}
\end{eqnarray*}
is the Riemann curvature two--form of the gravitational instanton metric.
Therefore $F$ is a self--dual Yang--Mills field
\[
F=*F,
\]
where $*$ is the Hodge operator on the gravitational instanton background.

The two--form $R_{ij}$ can be decomposed in terms of 
the Ricci tensor, Weyl tensor and Ricci scalar as
$
R_{ij}=W_{ijk}\Sigma_k+\Phi_{ijk}\Omega_k,
$
where $\Sigma_k$ and $\Omega_k$ are basis of SD (see \ref{two_forms})
and ASD two--forms
respectively. The coefficients $\Phi_{ijk}$ have nine components
corresponding to  the trace--free Ricci tensor.
The Bianchi identity $R_{ij}\wedge\Sigma_j=0$
gives $W_{ijj}=0$, so $W_{ijk}$ can be further decomposed into a self--dual
Weyl tensor (with five independent components), and the totally
skew part $\Lambda\varepsilon_{ijk}$, where $\Lambda$ is a multiple of the 
Ricci scalar.
In the self--dual vacuum case we have $\Phi_{ijk}=0, \Lambda=0$. 
Using the identities
\[
\Sigma_i\wedge\Sigma_j=2\delta_{ij}\mbox{vol}, \quad 
\mbox{Tr}({\bf t}_i{\bf t}_j)=-\frac{1}{2}\delta_{ij}, \quad
\varepsilon_{ijk}\varepsilon_{kpq}=\delta_{ip}\delta_{jq}-\delta_{iq}\delta_{jp}\]
we find $ \mbox{Tr}(F\wedge F)=-(1/2)|W|^2\mbox{vol}$, where $|W|^2=W_{ijk}W^{ijk}$.
\vskip5pt
In the case considered in this paper, where $P_j$ are given by the 
one--forms (\ref{connection}) we find (with $\cdot =d/dr$)
\begin{eqnarray*}
F&=&(\dot{f}_1 dr\wedge\eta_1+(f_1-f_2f_3)\;\eta_2\wedge \eta_3)\otimes {\bf t}_1\\
&+&(\dot{f}_2 dr\wedge\eta_2+(f_2-f_1f_3)\;\eta_3\wedge \eta_1)\otimes {\bf t}_2\\
&+&(\dot{f}_3 dr\wedge\eta_3+(f_3-f_1f_2)\;\eta_1\wedge \eta_2)
\otimes {\bf t}_3.
\end{eqnarray*}
The instanton number is  $k=-c_2$, where the Chern number is given by
\[
c_2=-\frac{1}{8\pi^2}\int_M \mbox{Tr}(F\wedge F).
\]
In our case 
\[
\mbox{Tr}(F\wedge F)=\frac{d}{dr}\Big(f_1 f_2 f_3-\frac{1}{2}(f_1^2+f_2^2+f_3^2)\Big)
dr\wedge \eta_1\wedge \eta_2\wedge \eta_3,
\]
and integration by parts gives 
$k_{TN}=1$ and $k_{AH}=2.$ In evaluating the $r$--integrals
we took into account that the radial direction is oppositely oriented
in the AH and the Taub--NUT cases. This is a consequence of a fact (carefully discussed
in \cite{AMS11} ) that the Taub--NUT metric is self-dual for the orientation 
which in the limit $\epsilon\rightarrow 0$ gives
the standard orientation on $\C^2$. The Atiyah--Hitchin manifold on the other hand
is self--dual for the orientation opposite to the complex orientations 
given by the underlying hyper--K\"ahler structure.
\section{Skyrmions from spin connection holonomy}
\label{section3}
The Skyrme model is a non--linear theory of pions in three space 
dimensions \cite{Skyrme_paper}. 
The model does not involve quarks, and is to be regarded as a low energy, effective theory of
QCD. A static
skyrmion is a map
\[
U:\R^3\rightarrow SU(2)
\]
satisfying the boundary conditions $U\rightarrow{\bf 1}$ as $|{\bf x}|\rightarrow \infty$.
The boundary conditions imply that $U$ extends to a one--point compactification
$S^3=\R^3\cup \{\infty\}$, and thus $U$ is partially classified
by its integer topological degree taking values in $\pi_3(S^3)$. In the Skyrme model
this topological degree is identified with the baryon number, which by continuity is conserved under time evolution. The non--linear field equations resulting from the Skyrme Lagrangian are not integrable and no explicit solutions are known. In contrast
to other soliton models, the Bogomolny bound is not saturated in the Skyrme case, and the energy is always greater than the baryon number. 

A good approximation of skyrmions is given by holonomy of $SU(2)$ instantons in $\R^4$, computed
along straight lines in one fixed direction \cite{AM89}. Choosing the lines to be parallel
to the $s=x^4$ axis gives the Atiyah--Manton ansatz
\[
U({\bf x})={\mathcal P}\exp{\Big(\int_{-\infty}^{\infty} 
A_4({\bf x}, s)ds\Big)}
\]
where $A_4$ is a component of the Yang--Mills instanton on  $\R^4$. The end points
of each line should be identified with the north--pole of the four sphere compactification
of $\R^4$. The boundary conditions at ${\bf x}\rightarrow \infty$ are then
satisfied as small circles on $S^4$ corresponding to straight lines 
shrink to a  point (Figure 1). Moreover the instanton number of $A$ is equal to the 
baryon number of the resulting skyrmion.
\begin{center}
\includegraphics[width=10cm,height=5cm,angle=0]{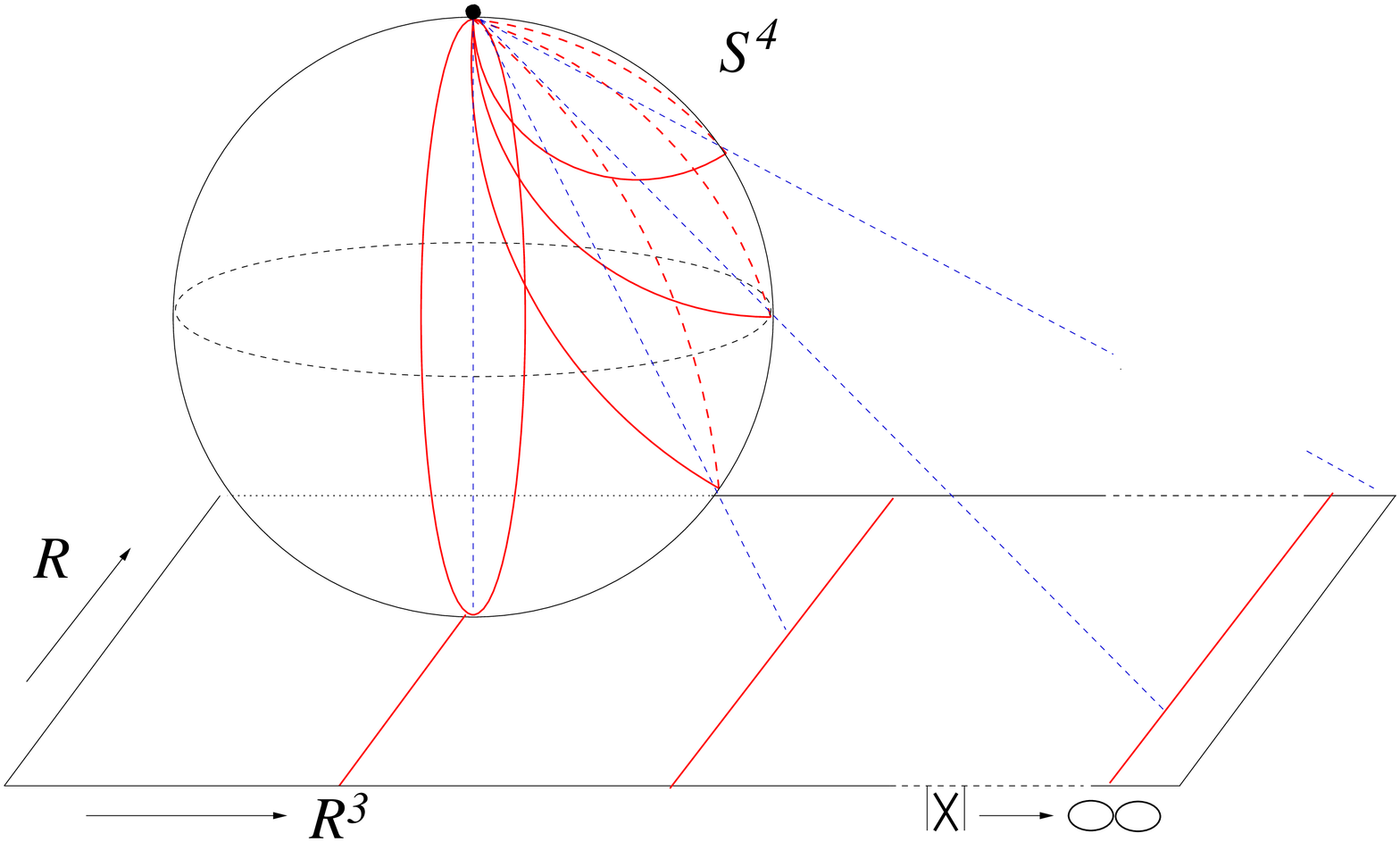}
\begin{center}
{{\bf Figure 1.} {\em Boundary conditions from the instanton ansatz.}}
\end{center}
\end{center}

In this section we shall adapt the Atiyah--Manton construction to gravitational 
instantons. While the underlying principle still applies, and holonomy of the spin connection along certain geodesics gives rise to a scalar group--valued field, the 
properties of the resulting skyrmion are very different. In particular
it is not defined on $\R^3$ which affects the boundary conditions. The topological degree
can still be found, but we shall not interpret it as the baryon number as this is given by
the signature of the underlying gravitational instanton. In particular the baryon number
is zero for the Taub--NUT skyrmion, but the Skyrme topological degree is not.

Let $K=K^a \p/\p x^a, a=0, \dots, 3$ be a vector field generating a 
one--parameter group of transformations of $M$ with orbits $\Gamma$. 
The holonomy of the gauge field
$A$ along $\Gamma$ arises from a solution to an ordinary differential equation
$K^aD_a \Psi=0$, where $D_a=\p_a+A_a$, and $\Psi=\Psi(s, x^j)$ 
takes its value in the
Lie group $SU(2)$. Here $K=\p/\p s$ and $x^j$ are the coordinates on the space
of orbits ${\mathcal B}$ of $K$ in $M$. If the trajectories $\Gamma$ are non--compact, then
one imposes the initial condition $\Psi(0, x^j)={\bf 1}$ at $s=-\infty$,
and sets the Skyrme field to be $U(x^j)=\lim_{s\rightarrow \infty}\Psi(s, x^j)$.
This gives
\be
\label{integral}
U={\mathcal P}\exp{(-\int_\Gamma A)},
\ee
where ${\mathcal P}$ denotes the $s$--ordering. In the case of
$\Gamma$ being a circle one breaks it up into an interval.

We now have to choose the curves $\Gamma$ along which the holonomy is to be calculated. We shall need an explicit parametrisation
of the one--forms $\eta_i$ in the metric (\ref{metric})
\be
\label{forms_eta}
\eta_2+i\eta_1=e^{-i\psi}(d\theta+i\sin{\theta} d\phi), \qquad
\eta_3=d\psi+\cos{\theta}d\phi,
\ee
where to cover $SU(2)=S^3$ in the Taub--NUT case we require the ranges
\[
0\leq\theta\leq\pi, \quad 0\leq\phi\leq 2\pi, \quad
0\leq\psi\leq 4\pi
\]
so that $\int \eta_1\wedge \eta_2\wedge \eta_3=-16\pi^2.$
 In the AH case take $0\leq\psi\leq 2\pi$, and make an identification
$(\theta, \phi, \psi)\cong (\pi-\theta, \phi+\pi, -\psi)$ so that $\int \eta_1\wedge \eta_2\wedge \eta_3=-4\pi^2.$

 One natural choice for $\Gamma$ is the family of 
the asymptotic circles with the $\psi$ 
coordinate varying, but this gives a trivial result 
as the resulting 
Skyrme field depends only on $r$, is Abelian, 
and its topological charge vanishes.
More generally, the holonomy should be calculated along the 
curves  which are orbits of a Killing vector  (or at least a conformal
Killing vector) as otherwise the space of orbits $\mathcal{B}$ of $K$ in $M$
does not admit a metric even up to scale. However a metric on $\mathcal{B}$ is necessary to compute the energy of the skyrmion.
This  rules out the asymptotic circles in the AH
case.  

We shall instead pick a left translation $SO(2)$ inside the isometry group 
$SO(3)$ (or its double cover $SU(2)$) of (\ref{metric}). Without lose 
of generality we can always choose the Euler 
angles in (\ref{forms_eta})
so that  the generator of this left translation is the right invariant vector
field $K=\p/\p\phi$. The $S^1$ fibres of ${\mathcal B}$ have no points in common and the resulting skyrmion can only be defined up to conjugation. However the preferred gauge has been fixed by choosing the $SO(3)$ or $SU(2)$  invariant 
frame (\ref{two_forms}) in which the spin connection components are proportional to the  left--invariant one forms, and the coefficients only depend on the radial coordinate.  This procedure is analogous to the one used by Atiyah and Sutcliffe \cite{AS05}.  The Yang--Mills connection resulting from our procedure  is given in the radial gauge $A_r=0$ as
\[
A=f_1(r)\eta_1\otimes {\bf t}_1+f_2(r)\eta_2\otimes {\bf t}_2+f_3(r)\eta_3\otimes {\bf t}_3.
\]
The residual gauge freedom $A\rightarrow \rho A \rho^{-1} - d\rho\, \rho^{-1}$, where $\rho=\rho(\theta, \psi, \phi)\in SU(2)$
can either be fixed by demanding regularity of the point $r=0$ (which is singled out as the fixed point of the isometry $K$) or by imposing the symmetry requirement
\[
{\mathcal L}_{R_i} A=0,\quad i=1, 2, 3
\] 
where the right--invariant Killing vector fields $R_i$ generate left--translations
and thus preserve the left--invariant one--forms $\eta_i$
i.e. ${\mathcal L}_{R_i}\eta_j=0$ and ${\mathcal L}$ denotes the Lie derivative.

The anti--symmetric matrix  $\nabla_a K_b$  has rank four at $r=0$, where the norm of $K$ vanishes.
Thus the  Killing vector has an isolated fixed point $r=0$, which is an anti--NUT in the terminology of \cite{GH}. Therefore some care needs to be taken  when constructing the metric
on the three--dimensional domain of the skyrmion - this will be done in Section \ref{section4}. 

Restricting the left--invariant forms $\eta_j$ to the $\phi$--circles
gives
\[
\eta_j={{n}_j}\;d\phi,
\]
where the unit vector ${\bf{n}}$ is given in the 
{\em unusual}  spherical polar coordinates $(\psi, \theta)$ 
by
\[
{\bf{n}}=(\cos{\psi}\sin{\theta}, \sin{\psi}\sin{\theta}, \cos{\theta}).
\]
 The integral (\ref{integral}) can be performed 
explicitly, as the component $A_\phi$ of the gauge field 
(\ref{gauge_field}) does not depend on $\phi$. 
This yields
\be
\label{final_skyrme}
U(r, \psi, \theta)=\exp{\Big(-i\pi\sum_{j=1}^3 {f_j(r){n}_j}\tau_j\Big)},
\ee
where the Pauli matrices $\tau_j$ are related to the generators of $\mathfrak{su}(2)$ by
${\bf t}_j=(i/2)\tau_j$. The topological charge
of the skyrmion is\footnote{In the special case of the usual hedgehog ansatz
where $f_1=f_2=f_3=F(r)$, and $0\leq\psi\leq 2\pi$ this formula reduces to the known expression
\[
B=-2\int_{r_0}^\infty \sin{(\pi F)}^2\;\frac{d F}{dr}dr.
\]
}
\be
\label{charge}
B=-\frac{1}{24\pi^2}\int_\mathcal{B} \mbox{Tr}((U^{-1}dU)^3)
\ee
\[
=-\frac{1}{2\pi}\int
({\frac{d{f}_1}{dr}f_2f_3\;{{n}_1}^2
+{f_1}\frac{d{f}_2}{dr}f_3\;{n}_2^2+{f_1}{f}_2\frac{d{f}_3}{dr}\;{{n}_3}^2})
\;\frac{\sin{(\pi\kappa)}^2}{\kappa^2}\sin{\theta}\; dr\; d\theta \;d\psi,
\]
where
$
\kappa=\sqrt{{f_1}^2{n}_1^2+{f_2}^2{{n}_2}^2+{f_3}^2{n}_3^2}.
$

Let us first consider
the Taub--NUT case, where $f_j$ are given by 
(\ref{taub_nut}). The field $U$ does not satisfy the boundary conditions
usually expected from a skyrmion, as $U(0)={\bf 1}$,  and 
$f_j\rightarrow (-1, -1, 1)$
as  $r\rightarrow \infty$. It nevertheless gives rise to a well defined constant
group element at $\infty$, as there $(f_1{n}_1, f_2 {n}_2, 
f_3{n}_3)$ tends to a unit vector 
and, setting ${\bf k}=(-{n}_1, -{n}_2, {n}_3)$, we get
\[
\lim_{r\rightarrow\infty} U=\cos{(-\pi)}{\bf 1}+i({\bf k}\cdot{\bf\tau})\sin{(-\pi)}=
-{\bf 1}.
\]
The functions $f_1$ and $f_2$ monotonically decrease and $f_3$ monotonically increases.
Thus - as the angle $\psi$ varies between $0$ and $4\pi$ - each element of the target space 
except $U=-{\bf 1}$ has exactly two pre-images
in the space of orbits on $\p/\p\phi$. Thus
the topological charge of the Taub--NUT skyrmion is 
\[
B_{TN}=2.
\]
This is confirmed by evaluating the integral (\ref{charge}).
We stress that the value of this topological charge is intimately related
to the period of the $\psi$ coordinate. The density does not depend
on $\psi$, so of the range is $\psi\in [0, k\pi]$, then $B_{TN}=k/2$.

 In the Atiyah--Hitchin case $f_j\rightarrow (-1, -1, 1)$ when 
$r\rightarrow \infty$, and the Skyrme field tends to a constant group
element $-{\bf 1}$ at infinity.  We do not expect the skyrmion 
to have a constant value at $r=\pi$, as this  corresponds to a 
bolt two--surface in the AH manifold. 
Formulae (\ref{AHsmall_r}) give
\[
U(r=\pi, \psi, \theta)=\exp{(3i\pi\cos{\psi}\sin{\theta}\;\tau_1)}.
\]
This skyrmion has a constant direction in $\mathfrak{su}(2)$ at the surface of the bolt,
with magnitude varying along its boundary.  

The degree of the Atiyah--Hitchin skyrmion is still well 
defined, but we have so far failed in calculating it 
by direct integration. Instead we shall use the method of counting pre-images.
Following \cite{AH, GM} we consider the parametrisation
of the radial functions $a_i(r)$ by elliptic integrals. 
Set
\begin{eqnarray*}
r&=&2 K(\sin(\beta/2)),\\
w_1&=&-r\frac{d r}{d\beta}\sin{\beta}-\frac{1}{2}r^2(1+\cos{\beta}),\\
w_2&=&-r\frac{d r}{d\beta}\sin{\beta},\\
w_3&=&-r\frac{d r}{d\beta}\sin{\beta}+\frac{1}{2}r^2(1-\cos{\beta})
\end{eqnarray*}
where $K$ is the elliptic integral
\[
K(k)=\int_0^{\pi/2}\frac{1}{\sqrt{1-k^2\sin{\tau}^2}}d\tau
\]
so that when $\beta\in[0, \pi)$, then $r(\beta)\in [\pi, \infty)$ is a monotonically increasing function.  
The radial functions in the AH metric are
\[
a_1=\sqrt{\frac{w_2w_3}{w_1}}, \quad a_2=\sqrt{\frac{w_1w_3}{w_2}},\quad
a_3=-\sqrt{\frac{w_1w_2}{w_3}},
\]
and we find the spin connection coefficients (\ref{connection}) to be
\begin{eqnarray*}
f_1&=&\frac{1}{2}\Big(\frac{a_3}{a_2}+\frac{a_2}{a_3}- \frac{{a_1}^2}{a_2a_3}  \Big)-\frac{r}{a_2}  \frac{d a_1}{dr},\\
f_2&=&\frac{1}{2}\Big(\frac{a_3}{a_1}+\frac{a_1}{a_3}-\frac{{a_2}^2}{a_1a_3}\Big)-\frac{r}{a_2}  \frac{d a_2}{dr},\\
f_3&=&\frac{1}{2}\Big(\frac{a_1}{a_2}+\frac{a_2}{a_1}-\frac{{a_3}^2}{a_1a_2}\Big)-\frac{r}{a_2}  \frac{d a_3}{dr}.
\end{eqnarray*}
The graphs of $r(\beta), a_i(\beta), f_i(\beta)$ are shown on Figure 2 (the graphs provided in 
\cite{GM} give
$a_i$ as functions of $r$.).
\begin{center}
\includegraphics[width=6cm,height=8cm,angle=0]{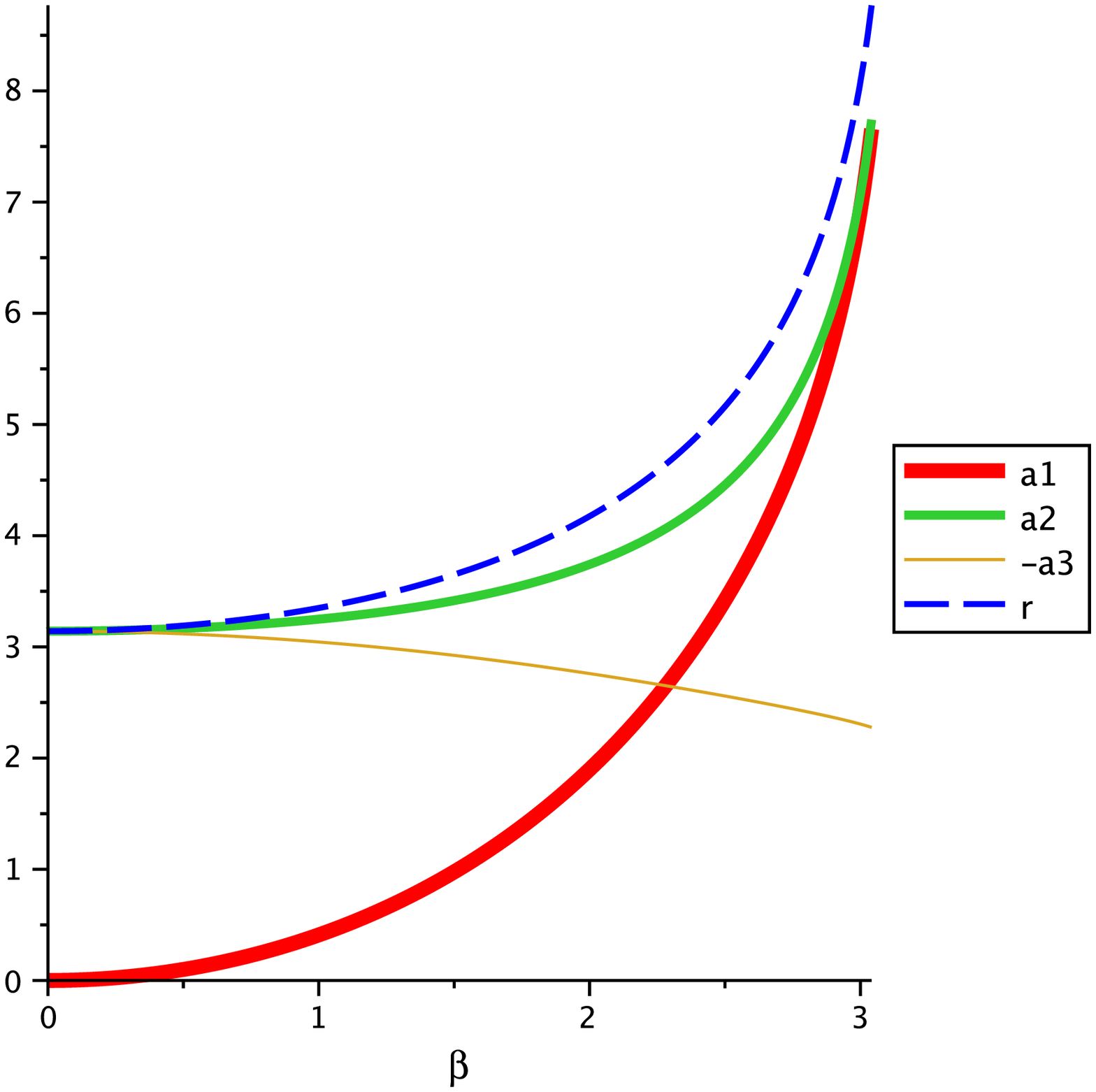}
\includegraphics[width=6cm,height=8cm,angle=0]{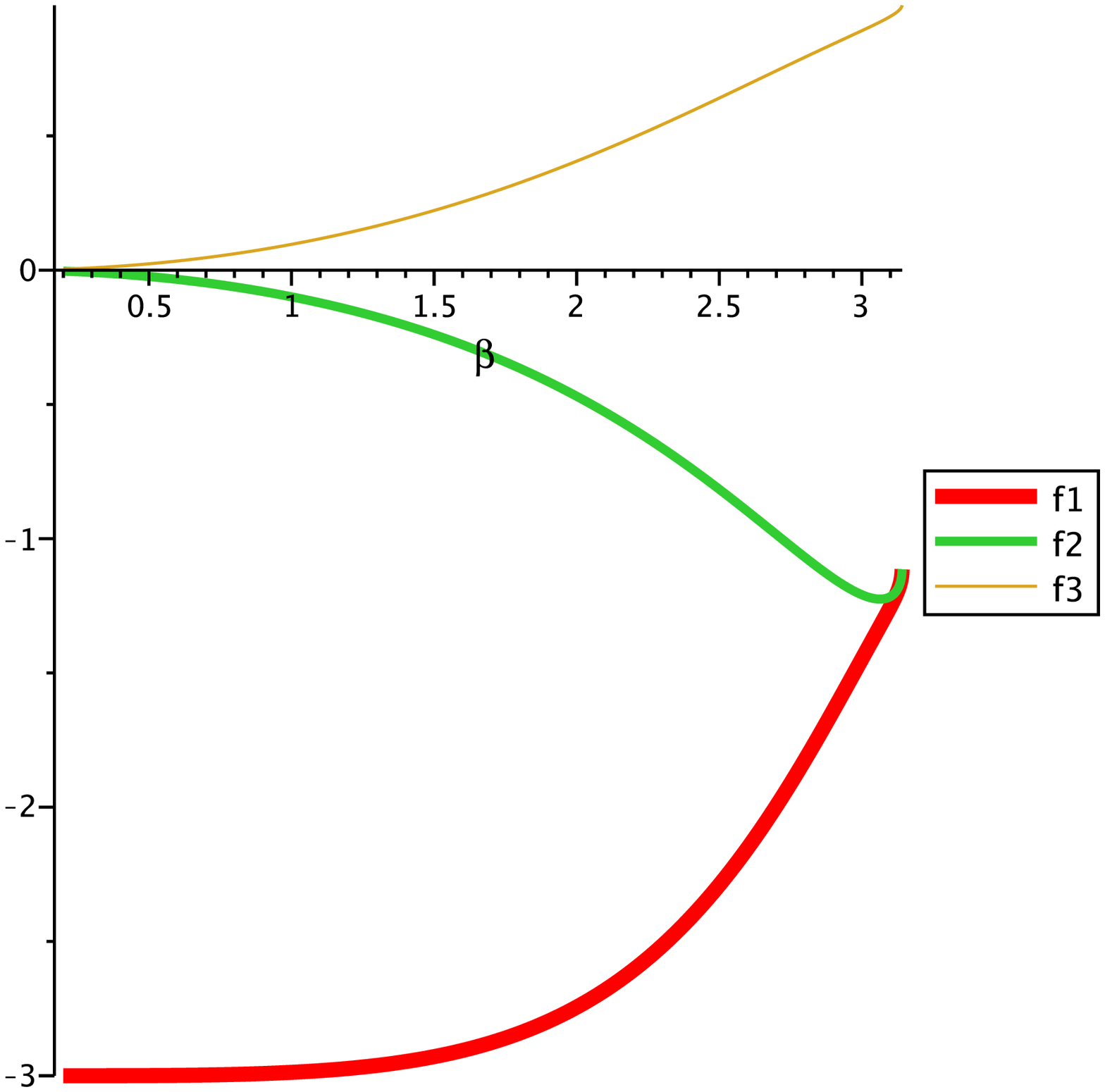}
\begin{center}
{ {\bf Figure 2.}  {\em Radial  functions $a_1, a_2, a_3$ and $r$ (1st graph)
and the spin connection coefficients (2nd graph) as functions of $\beta\in [0, \pi)$.}}
\end{center}
\end{center}
The functions $f_1$ and $f_3$ are monotonically increasing,
and the function $f_2$ appears to have a local minimum  near $\beta=3.0635$. Thus each point
on $SU(2)$ except $U=-{\bf 1}$ has exactly one pre-image in ${\mathcal B}$, and the topological charge of the Atiyah--Hitchin skyrmion is $B_{AH}=1$. 
\section{Geometry of the Taub--NUT skyrmion}
\label{section4}

 In both AH and Taub--NUT cases
the resulting skyrmion is defined on the space of orbits 
${\mathcal B}$ of the isometry
$\phi\rightarrow\phi+{const}$ in $(M, g)$. This space
has a natural conformal metric induced by (\ref{metric}). To find it, 
perform the standard Kaluza--Klein reduction on $\phi$, simply  by completing 
the square. Set
\[
\Omega^2={a_1}^2\cos{\psi}^2\sin{\theta}^2+{a_2}^2\sin{\psi}^2\sin{\theta}^2+
{a_3}^2\cos{\theta}^2.
\]
Then
\[
g=h+\Omega^2(d\phi+\Omega^{-2}\omega)^2,
\]
where $(h, \omega)$ are a metric and a one form respectively on the space of 
orbits  of $\p/\p\phi$ given by
\[
h=f^2 dr^2+ ({a_1}^2\sin{\psi}^2+{a_2}^2\cos{\psi}^2) d\theta^2 +{a_3}^2d\psi^2
-\Omega^{-2}\omega^2,
\]
\[
\omega=({a_2}^2-{a_1}^2)\sin{\psi}\cos{\psi}\sin{\theta} d\theta
+{a_3}^2\cos{\theta}d\psi.
\]
The metric $h$ is only defined up to scale on the space of orbits, and we can choose this scale freely.

One choice of the conformal factor which takes into account the range 
of the angular coordinate $\theta$
in the Taub--NUT case is\footnote{This conformal metric also admits a Weyl
connection such that the Einstein--Weyl equations hold on ${\mathcal B}$.
This is true for any conformal structure on the space of orbits
of an isometry in a Riemannian manifold with self--dual Weyl curvature -- see e.g. \cite{Dunajski_book}}
\[
h_{\mathcal B}=\frac{1}{{a_1}^2\sin{\theta}^2}\;h=g_{\HH^2}+R^2\;d\psi^2
\]
This metric is defined on a trivial circle bundle over the hyperbolic plane.
The vector $\p/\p\psi$ is an isometry of $h_{\mathcal B}$ which is a consequence
of the fact that the right translations
$\psi\rightarrow \psi+const$ of the Taub--NUT space
commute\footnote{This would not be the case for the AH metric, where the domain of the resulting skyrmion does not admit any isometries.}
 with the left translations $\phi\rightarrow \phi+const$.
In the upper half plane model where $x=r\cos{\theta}, y=r\sin{\theta}$
the hyperbolic metric $g_{\HH^2}$
and the varying radius $2R$ (as $\psi$ is between $0$ and $4\pi$) of 
the $S^1$ fibres are given by
\[
g_{\HH^2}=\frac{dx^2+dy^2}{y^2}, 
\quad R^2=\frac{1}{(a_1/a_3)^2\sin{\theta}^2+\cos{\theta}^2}=
\frac{x^2+y^2}{y^2({\mu}\sqrt{x^2+y^2}+1)^2+x^2},
\]
where $\mu=\epsilon/m$. The radius of the circles tends to two on the real line boundary
of the upper half--plane, and shrinks away from the boundary. The metric 
is complete, as the radius does not vanish anywhere
on ${\mathcal B}$.  The density of the resulting skyrmion (\ref{final_skyrme}) attains its maximum
at the $y$--axis in the upper half--plane model, where it is given by
\[
\frac{\pi\mu y(\mu y+2)}{(\mu y+1)^2}
\sin{\Big(\frac{\pi \mu y}{\mu y+1}\Big)}^2.
\]
The location of the maximum is a root of the transcendental
equation $\tan{\hat{y}}=\hat{y}^{-1}(\pi^2-\hat{y}^2)$, where
$\mu y=\pi/\hat{y}-1$. The approximate solution is $y=5/(4\mu)$, and
the resulting maximal density is approximately $1.22$ for all values
of the parameter $\mu$. The skyrmion density is  independent on $\psi$.
\begin{center}
\includegraphics[width=5cm,height=5cm,angle=0]{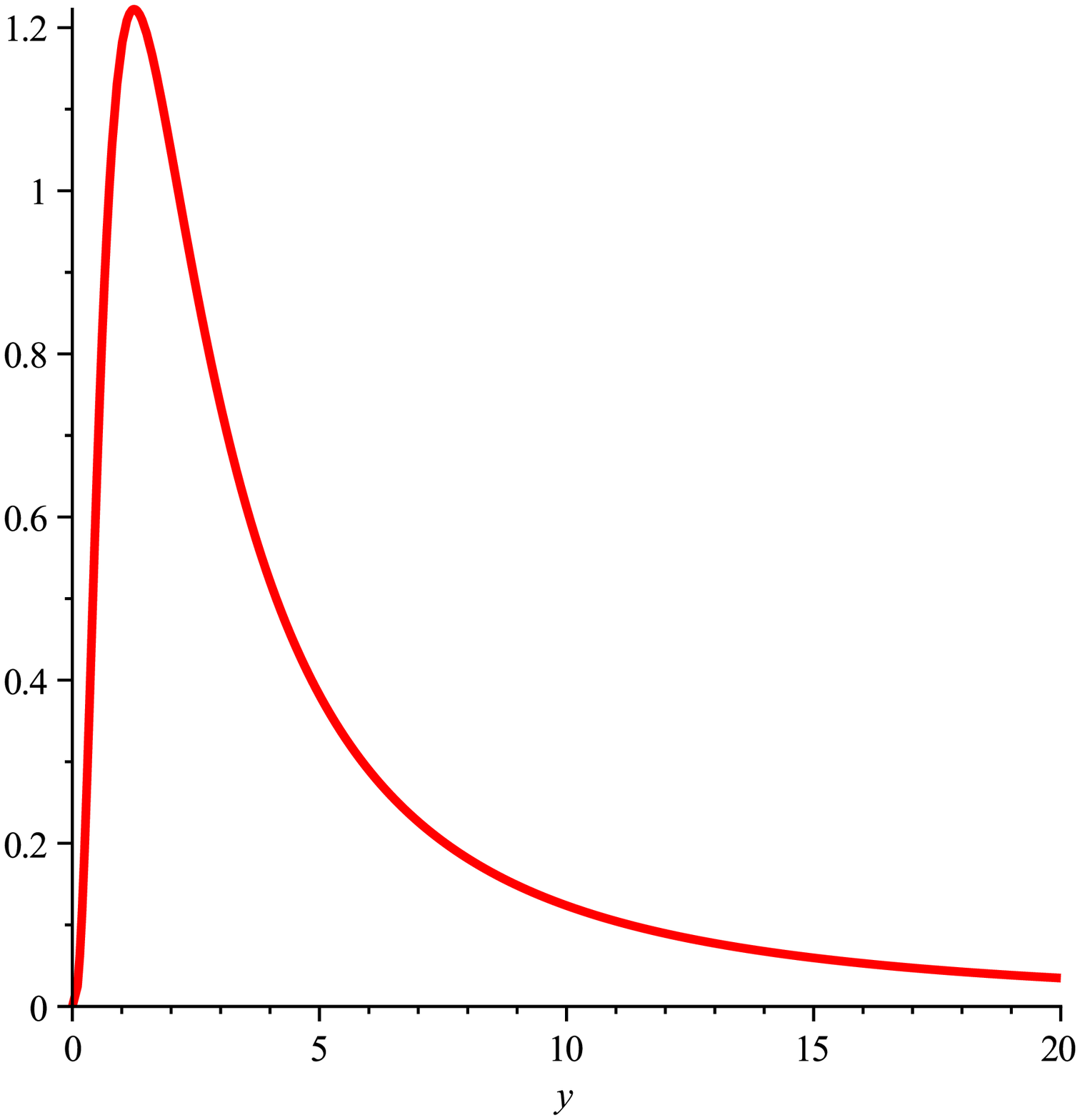}
\includegraphics[width=8cm,height=8cm,angle=0]{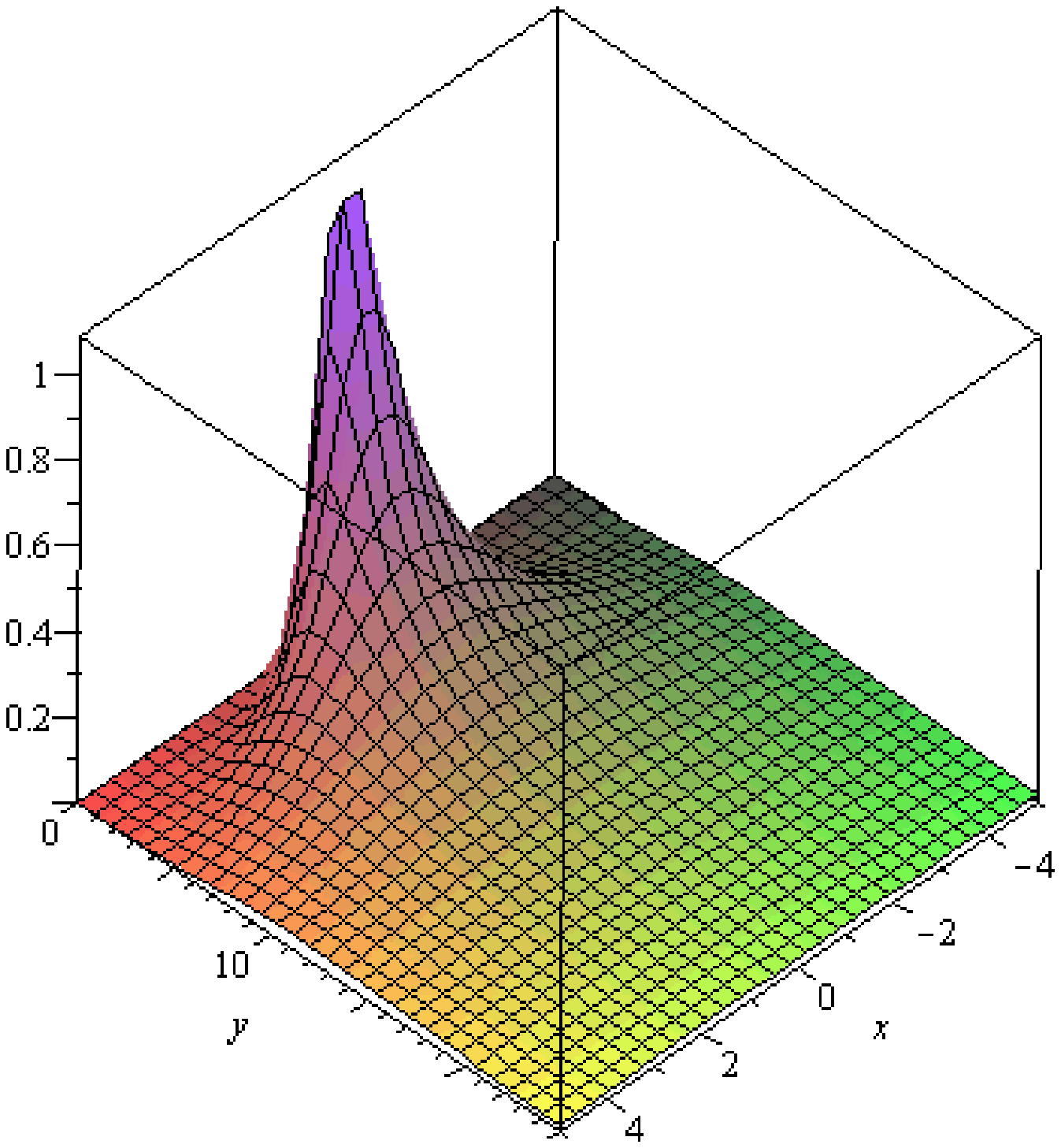}
\begin{center}
{{\bf Figure 3a.} {\em Skyrmion density on the upper half plane.}}
\end{center}
\end{center}
In the disc model of the hyperbolic space,
the boundary of ${\mathcal{B}}$
is a flat torus. Let the map
$\DD\rightarrow \HH^2$ be given by
\[
x+iy=\frac{z-i}{iz-1},
\]
where $|z|<1$. The radius of fibers of $S^1\rightarrow 
{\mathcal B}\rightarrow \DD$ is discontinuous at the point
$z=-i$
corresponding to $y=\infty$ on the boundary (Figure 3b). 
\begin{center}
\includegraphics[width=5cm,height=5cm,angle=0]{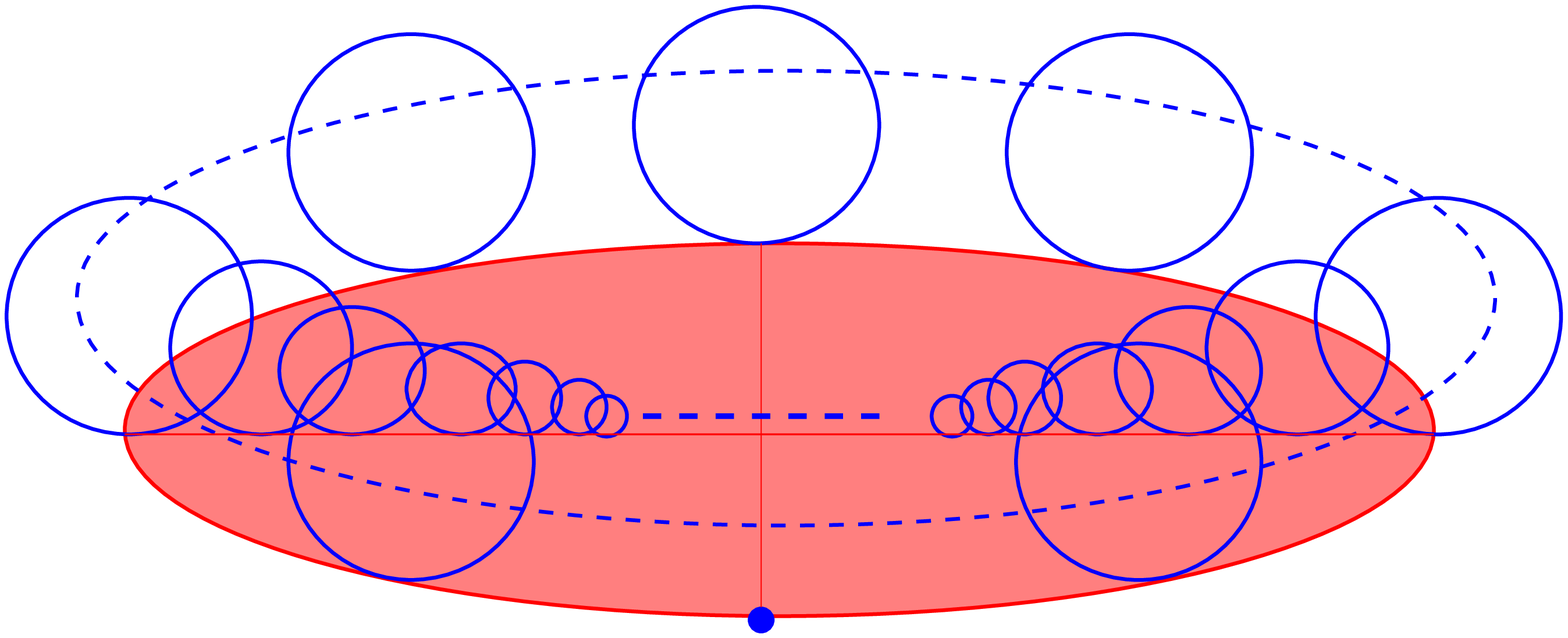}
\begin{center}
{{\bf Figure 3b.} {\em Circle fibration over the Poincare disc, with 
shrinking fibers.}}
\end{center}
\end{center}
The Ricci scalar of
$h_{\mathcal B}$ is also discontinuous on the boundary, and equals
$-2$ if $|z|=1$ and $z\neq -i$, and $-6$ at $z=-i$ (Figure 4). At the centre of the disc
the Ricci scalar depends on $\mu$ and is given by $-2(3\mu^2+3\mu+1)/(\mu+1)^2$.
This does not cause a  problem as the point $z=-i$ is not a part of the 
manifold ${\mathcal B}$. 
\begin{center}
\includegraphics[width=6cm,height=8cm,angle=0]{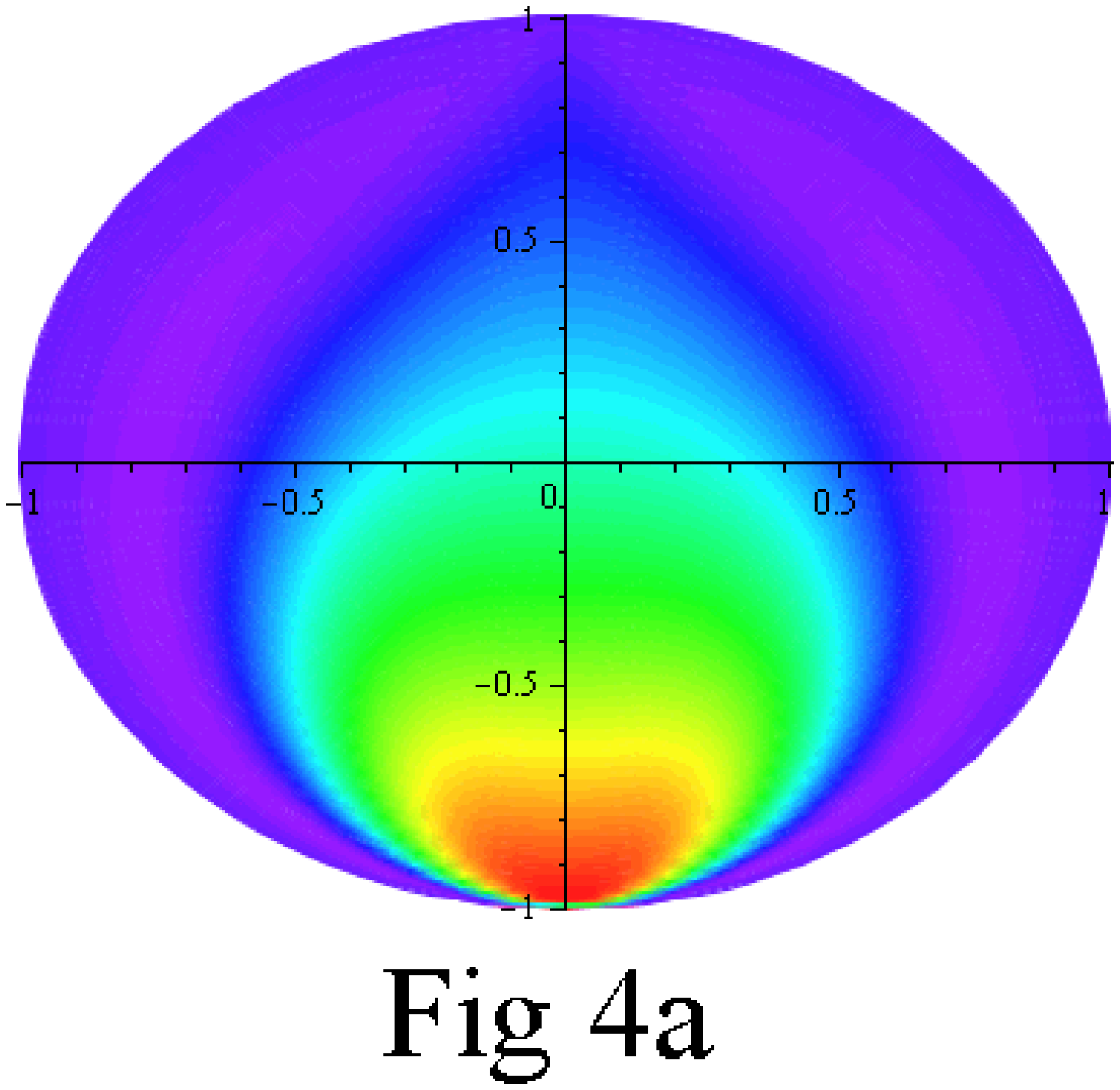}
\includegraphics[width=6cm,height=8cm,angle=0]{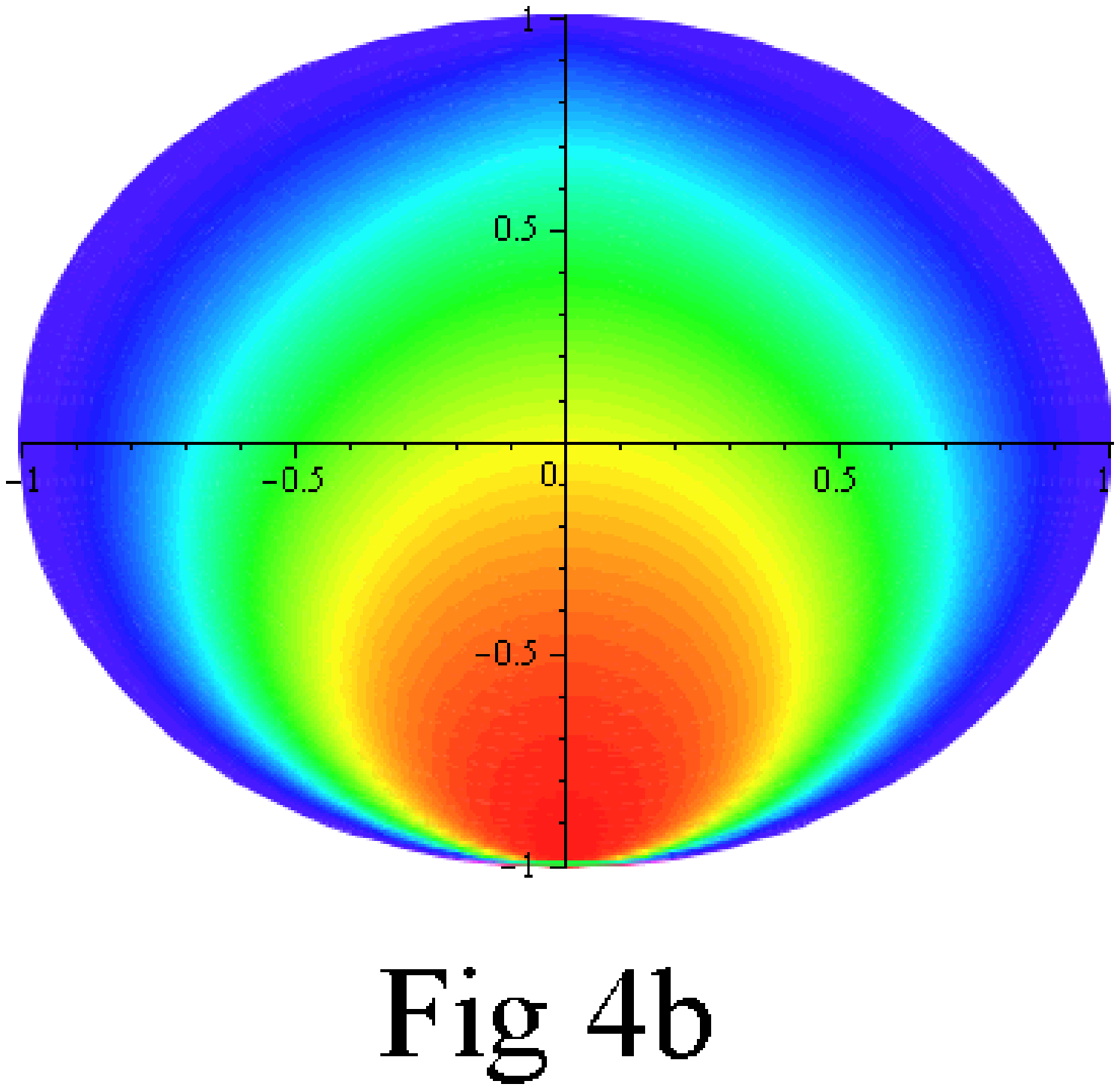}
\begin{center}
{{\bf Figure 4.} {\em Density plots of the Ricci scalar (Fig. 4a) of  
$({\mathcal B}, h_{\mathcal B})$  and the radii of the circles (Fig. 4b)
in the fibration $S^1\rightarrow \DD$ for $\mu=1$.}}
\end{center}
\end{center}
The density of the Taub--NUT skyrmion peaks on the diameter joining $z=-i$ 
and $z=i$. The skyrmion is located on this diameter at
$z=i(4\mu-5)/(4\mu+5)$. (Figure 5).
At the centre of the disc the skyrmion varies along the fibres according to
\[
U=\exp{(i\pi\mu/(\mu+1)(\cos{\psi}\tau_1+\sin{\psi}\tau_2)}).
\]
The point
$r=0$ corresponds to the point $z=i$ on the boundary of the disc, where $U={\bf 1}$.
Note that this point has been removed from the domain of the skyrmion, as it is the fixed
point (in four dimensions) of the isometry used to obtain the 
quotient ${\mathcal B}$.
The boundary conditions  $r\rightarrow \infty$ translate to $U=-{\bf 1}$ at the point $z=-i$ for all $\psi$.
\begin{center}
\includegraphics[width=15cm,height=15cm,angle=0]{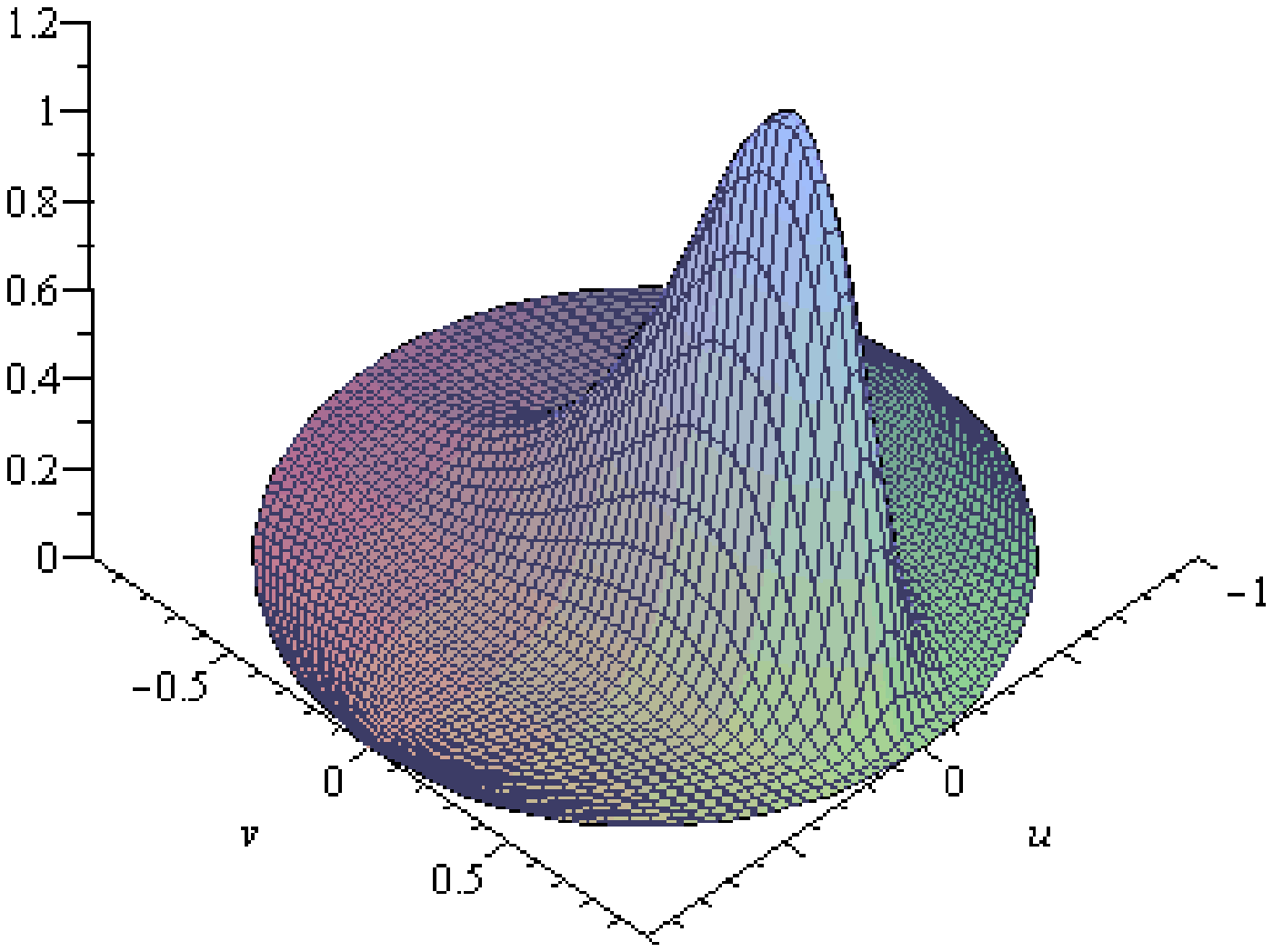}
\begin{center}
{{\bf Figure 5.} {\em Density of the Taub--NUT skyrmion 
in the Poincare disc model with $z=u+iv, \mu=4$}.}
\end{center}
\end{center}
\section{Further remarks}
We  have constructed $SU(2)$--valued  Skyrme fields from a holonomy 
of a non--flat $\mathfrak{su}(2)$ spin connection 
on $\spp_+$ corresponding to ALF gravitational instantons. The holonomy is 
calculated along orbits of an isometry generating a one--parameter
subgroup $SO(2)$ of the full isometry group. In case of Taub--NUT,
the Skyrme field carries a non--zero topological charge. This rules out
the interpretation of the charge as the baryon number -- which vanishes -- 
but opens up a possibility of assigning other integral charges 
to particles in the AMS model. A lepton number is an obvious 
candidate\footnote{It may be that some combination of the Skyrme charge, and the Euler number and the signature corresponds to the lepton number. An identification of the lepton number with
$(B-1)$ is consistent with the AMS proposal, but possibly too naive.}, 
as no proposal of its topological interpretation has been put forward
in \cite{AMS11}. 

A computation of the Skyrme field can  be carried over for other 
gravitational instantons. To do it for the Fubini--Study metric on $\CP^2$,
one needs to express it in the Bouchiat--Gibbons form
\cite{BG} adapted to the $SO(3)$
(rather than $U(2)\subset SU(3))$ action. This, with $r\in [0, \pi/2]$, leads
to a connection (\ref{connection}) with
\begin{eqnarray*}
f_1&=&\frac{2\sin{(r+\pi/2)}\cos{r}+\cos{r}^2}{\sin{(r+\pi/2)}},\\
f_2&=&\frac{2\cos{(r/2+\pi/4)}^2+2\sin{r}\cos{(r/2+\pi/4)}^2-\cos{r}^2}
{2\sin{r}\cos{(r/2+\pi/4)}},\\ 
f_3&=&-\frac{2\cos{(r/2+\pi/4)}^2+2\sin{r}\sin{(r/2+\pi/4)}^2+\cos{r}^2-2}
{2\sin{r} \sin{(r/2+\pi/2)}}.
\end{eqnarray*}
The instanton number of the corresponding gauge field is fractional
$k_{\CP^2}=9/2$, which reflects the fact that $\CP^2$ does not admit a spin
structure, and the gauge field resulting from the spin connection is not 
globally defined. The resulting Skyrme field  behaves similarly to the AH case, as $(f_1, f_2, f_3)$ equals $(3, 0,  0)$  at $r=0$, and $(0, 0, 0)$ at 
$r=\pi/2$ . While the number and the position of nuts and bolts might
depend on the choice of the isometry --  $\CP^2$ has three nuts for
$\p/\p\phi$ and a nut and a bolt for $\p/\p\psi$ --
the total number of nuts and bolts is constrained by a topological
equality \cite{GH}
\[
\sum \mbox{nut}+\sum\chi(\mbox{bolt})=\chi(M)
\]
and the RHS is equal to $3$ for $\CP^2$.

 We can also compute the Skyrme field for the Euclidean 
Schwarzschild metric. This gravitational instanton does not appear
in \cite{AMS11}, as its curvature is not self--dual. It is nevertheless
possible that it can be used as a model for the 
neutrino in place of  $S^4$ - the case for self--duality 
of the underlying Riemannian manifolds
was not overwhelmingly strong in \cite{AMS11}. The topology
of the Euclidean Schwarzschild is $\R^2\times S^2$, 
with boundary $S^1\times S^2$.  Now there are two Yang--Mills fields constructed out of self--dual and anti--self--dual spin connections, and the Euler and Pontriagin numbers are
linear combinations of the two instanton numbers.
The Schwarzschild manifold has signature zero, 
which is compatible with AMS interpretation.
The metric is asymptotically flat, and the asymptotic circle fibration
is trivial. Thus there is no associated electric charge.

\subsubsection*{Acknowledgements}
I thank Gary Gibbons, Nick Manton and Prim Plansangkate for useful discussions. I also thank the anonymous referees
for their comments which led to several improvements in the manuscript.

\end{document}

%% file: papers_macros.tex
\newcommand{\C}{\mathbb{C}}
\newcommand{\CP}{\mathbb{CP}}
\newcommand{\HH}{\mathbb{H}}

\newcommand{\R}{\mathbb{R}}

\newcommand{\DD}{\mathbb {D}}

\def\p{\partial}

\newcommand{\bea}{\begin{eqnarray}}
\newcommand{\eea}{\end{eqnarray}}

\def\be{\begin{equation}}
\def\ee{\end{equation}}

\def\p{\partial}

\newcommand{\spp}{\mathbb{S}}